\documentclass[12pt,a4paper]{article}

\usepackage[scale={.75,.8}]{geometry} % Wide page
\usepackage{pstricks}
\usepackage[dvips]{graphicx}
\usepackage[latin2]{inputenc}
\usepackage{epic}
\usepackage{latexsym}
\usepackage{epsf}
\usepackage{epsfig}
\usepackage{amssymb,amsmath}
\usepackage{cite} %gives 1--4 instead of 1,2,3,4 in refs. 

\newcommand{\eref}[1]{eq.\ (\ref{e.#1})} 
\newcommand{\erefn}[1]{ (\ref{e.#1})}
 
\newcommand{\aref}[1]{\ref{a.#1}}
\newcommand{\sref}[1]{Section \ref{s.#1}}
\newcommand{\cref}[1]{Chapter \ref{c.#1}}

\def\nn{\nonumber \\}  
\newcommand{\nl}{& \nonumber \\ &}
\newcommand{\bnl}{\right . & \nonumber \\ & \left .}

\def\beq{\begin{equation}} 
\def\eeq{\end{equation}} 
\def\bea{\begin{eqnarray}}  
\def\eea{\end{eqnarray}}  
\newcommand{\bal}{\begin{align}}
\newcommand{\eal}{\end{align}}   
\def\ba{\begin{array}}  
\def\ea{\end{array}}   
\def\bi{\begin{itemize}}  
\def\ei{\end{itemize}}  
\def\ben{\begin{enumerate}}  
\def\een{\end{enumerate}}  
\def\beq{\begin{equation}}  
\def\eeq{\end{equation}}  
\def\bc{\begin{center}}
\def\ec{\end{center}} 
 \def\bt{\begin{table}}
\def\et{\end{table}}  
 \def\btb{\begin{tabular}}
\def\etb{\end{tabular}}  

\newcommand{\bvec}{\left ( \ba{c}}
\newcommand{\evec}{\ea \right )}

\def\cv{{\mathcal V}}
   
\def\cl{{\mathcal L}}

\def\co{{\mathcal O}}

\def\tev{\, {\rm TeV}}

\def\mass2{mass${}^2$}

\def\tev{\, {\rm TeV}}

\def\mass2{mass${}^2$}

\newcommand{\mpl}{M_{\mathrm{Pl}}}  
  
\newcommand{\mkk}{M_{\mathrm{KK}}}  
\newcommand{\ads}{$\bf \mathrm{AdS}_5$}

\def\ra{\rangle}
\def\la{\langle}  

\def\pa{\partial}

\newcommand{\tr}{\rm Tr}

\newcommand{\ti}{\tilde}  
\def\hc{{\rm h.c.}}

\def\ov{\overline}

\begin{document}

\pagestyle{empty}
\begin{flushright}
CERN-PH-TH/2007-038\\ 

{\bf \today}
\end{flushright}
\vspace*{5mm}
\begin{center}

\renewcommand{\thefootnote}{\fnsymbol{footnote}}

{\large {\bf Electroweak observables in a general 5D background}} \\ 
\vspace*{1cm}
{\bf Antonio~Delgado$^{a}$\footnote{Email:antonio.delgado@cern.ch} ,}
{\bf Adam~Falkowski$^{a,b}$}\footnote{Email:adam.falkowski@cern.ch}

\vspace{0.5cm}

$^{a}$CERN Theory Division, CH-1211 Geneva 23, Switzerland\\
$^{b}$Institute of Theoretical Physics, Warsaw University, \\ Ho\.za 69, 00-681 Warsaw, Poland

\vspace*{1.7cm}
\begin{abstract}

\emph{Warped} extra dimensions provide a new playground to study electroweak brea-king and the nature of the higgs field. In this paper we reanalyze the electroweak observables in theories with one extra dimension and a completely general  warp factor. We demonstrate that, regardless of the 5D background, $SU(2)_R$ is needed in order to avoid excessive contributions to the $T$ parameter. For higgsless theories cancellations between different contributions to the $S$ parameter are needed.

\end{abstract}
\end{center}
\vspace*{5mm}
\noindent 

\vspace*{1.0cm}
\date{\today}

%\noindent PACS numbers: \ldots

%\rule[.1in]{16.5cm}{.002in}
\vspace*{0.2cm}
 
\vfill\eject
\newpage

\setcounter{page}{1}
\pagestyle{plain}

\renewcommand{\thefootnote}{\arabic{footnote}}
\setcounter{footnote}{0}

%%%%%%%%%%%%%%%%%%%%%%%%%%%%%%%%%%%%%%%%%%%%%%%%
\section{Introduction}
%%%%%%%%%%%%%%%%%%%%%%%%%%%%%%%%%%%%%%%%%%%%%%%%%
The \emph{precise} way the electroweak symmetry  is broken is the last aspect left to be discovered to have a complete description of particle physics based on the group $SU(3)_c\times SU(2)_L \times U(1)_Y$. 
It is therefore one of the key aspects to be studied at the Large Hadron Collider (LHC).
In the standard model (SM) one single fundamental scalar field, the higgs, is responsible for the breaking but its nature and the reason for that scalar to break the electroweak symmetry is not explained. 
This is the main reason to go beyond the SM. 
The two standard approaches are either to explain why a fundamental scalar is not unstable under radiative corrections (SUSY) or to construct models for dynamical electroweak breaking (technicolor or a pseudo-goldstone higgs). 

Just before LEP-II the former approach was favoured due to the problems of models with dynamical breaking with the electroweak observables \cite{Peskin:1991sw}. But after LEP-II and the \emph{no}-discovery of the higgs or any supersymmetric particle there has been a revival of interest in models with  a composite higgs or even without a higgs. The progress in model-building has been possible thanks to the formulation of strongly coupled theories as  models with a \emph{warped} extra dimension \cite{rasu}. 

Theories formulated in 5D with the geometry of a slice of \ads \, are in close relation\cite{adscft} with 4D strongly coupled, approximately conformal  theories which eventually confine and have also a fundamental sector that couples weakly to these bound states. The bound states map to the KK modes in the 5D theory whereas the fundamental fields correspond to the zero modes of the 5D theory. One of the main advantages of this formulation is that in 5D one can carry out calculations and have observable quantities under control. 

In this spirit, the electroweak sector of the SM can be formulated as a 5D warped model with two boundaries, one being the UV and the second one the IR brane. Gauge bosons and chiral fermions are free to move in the bulk of the extra dimension. The breaking of the electroweak symmetry occurs via a localized vev on the IR brane, which means that the higgs does not propagate into the bulk \cite{agdema}. The hierarchy problem is solved because the higgs is interpreted as a composite object beyond the IR scale and this scale is set to be around a TeV. 
One can even send its vev to infinity and have a theory without a higgs (light scalar resonance) in the spectrum\cite{Csaki:2003zu}. 
A third possibility is to suppose that the higgs is a 5D field but embedded in a gauge multiplet,  so that  its mass is protected\cite{agcopo}. 
This gauge symmetry is broken via boundary conditions and the higgs field arises as the lowest mode of the fifth component of the gauge field. 
This is mapped in the 4D picture to a theory where the higgs is a pseudo goldstone boson corresponding to a spontaneously broken global symmetry of the strongly coupled sector. 

All these theories have been analyzed and  different electroweak parameters have been calculated \cite{agdema,cserte,agco,Cacciapaglia:2004jz,caposa2}. 
%In general, custodial symmetry in the bulk is needed to prevent large contributions to the $T$ parameter.  The $S$ parameter is under control whenever there is a higgs.  In the case of higgsless theories some cancelation with fermionic contributions is required. 
In this paper we are going to revisit the first two scenarios, those where the breaking is localized on the IR brane,  but for a general warped metric and not only the one of \ads. 
On the 4D side, this corresponds to departure from an approximate conformal symmetry of the strongly coupled sector. 
We calculate tree-level contributions to the electroweak parameters both in the case with a higgs localized on the IR brane and in the higgsless theory  (we leave the study of  a pseudo-goldstone higgs for future publications).
We present general formulas for the spectrum of gauge bosons %and fermions, 
and their contributions to $S$, $T$ and $Zb\bar{b}$ couplings. 
The main conclusions is that, also in general backgrounds, $SU(2)_R$ is needed to cancel large contributions to $T$. The $S$ parameter is under control for the case with a higgs, while in the higgsless case cancellations are necessary to happen independent of the background metric.

The paper is organized as follows, in section 2 different general formulas for the spectrum and the matching conditions between the 5D and the 4D theories are given, in section 3 the SM is studied leading to a large contribution to the $T$ parameter, $SU(2)_R$ is introduced in section 4, higgsless models are studied in section 5 and finally our conclusions are presented in section 6. Some technical details are given in the Appendix.

%%%%%%%%%%%%%%%%%%%%%%%%%%%%%%%%%%%%%%%%%%%%%%%%%
\section{Tools}
\label{s.f} 
%%%%%%%%%%%%%%%%%%%%%%%%%%%%%%%%%%%%%%%%%%%%%%%%%

In this section we present the formalism we employ in order to derive electroweak precision constraints on 5D gauge theories in warped backgrounds.   
We choose to work in the KK picture.\footnote{Another approach, so-called holographic \cite{bapora}, consists in integrating out the bulk degrees of freedom and writing down an effective action for the UV brane degrees of freedom. Physical results, of course, do not depend on the approach.} 
We diagonalize the 5D action in the KK basis in the presence of electroweak breaking on the IR brane.
This approach is conceptually clear.
Moreover, derivation of the tree-level effective action for the  SM fields is simplified,    
as the zero mode fields do not mix the with the heavy modes. 
Thus, the gauge boson masses and the vertex corrections can be read off directly from the KK diagonalized  5D action. 
This information allows, in particular, to determine the oblique S and T parameters, which typically encode the most stringent bounds on the model.  
Determination of four-fermion operators  in the effective theory still requires computing diagrams with the heavy KK mode  exchange, but in most situations those do not impose additional constraints.  

Below we introduce the techniques that allow to perform the KK diagonalization and the integrating-out procedure for an arbitrary warped background. 
We also review the basic facts concerning the parametrization of physics beyond the SM by dimension six operators.

%**********************************************************************
\subsection{Kaluza-Klein expansion in general warped backgrounds}
%***********************************************************************

We study 5D gauge theories with the fifth dimension being an interval, $x_5 \in [0,L]$.
The gravitational background is described by the line element    
\beq
\label{e.wb}
ds^2 = a^2(x_5) \eta_{\mu\nu} dx^\mu dx^\nu - dx_5^2 \, .
\eeq  
with the warp factor $a(x_5)$.  We fix $a(0) = 1$. 
The choice $a(x_5) = 1$ corresponds to 5D flat spacetime, while $a(x_5) = e^{- k x_5}$ corresponds to \ads. 
For most of the subsequent discussion we do not specify the warp factor.
We only assume that it is a monotonic and non-increasing function, so that it makes sense to define a UV brane at  $x_5 = 0$ and an IR brane at $x_5 = L$, where the value of the warp factor is $a(L) \equiv a_L \leq 1$.

Consider the quadratic action for a 5D gauge field propagating in a warped background
\beq
S_5 = \int d^4 x \int_0^L dx_5  \left \{ -  {1\over 4} (\pa_{\mu}A_\nu - \pa_{\nu} A_\mu)^2  + {1 \over 2} a^2(x_5) (\pa_5 A_\mu )^2 
%+  {1 \over 2} L \ti m_0^2 A_\mu^2  \delta(0)  
+  {1 \over 2} L \ti m_L^2 A_\mu^2  \delta(L)  \right \}
\eeq
The boundary mass term represents the effect of the boundary higgs field vev. 
We expand the 5D gauge field in the KK basis 
\beq
\label{e.kkp} % KK profiles 
A_\mu(x,x_5) = \sum_n f_n(x_5) A_{\mu,n}(x)
%\quad   
%A_5(x,x_5) = \sum_n {\pa_5 f_n(x_5) \over m_n(h)} G_n(x)
\eeq
and choose the profiles such that the quadratic action, {\it in the presence of the boundary mass term},  can be rewritten as a 4D action diagonal in $n$: 
\beq
S_5 = \int d^4 x  \sum_n \left \{ -  {1\over 4} (\pa_{\mu}A_{\nu,n} - \pa_{\nu} A_{\mu,n})^2  + {1 \over 2} m_n^2 (A_{\mu,n})^2   \right \} \, .
\eeq
In this way, any mixing between a possible massless mode and the heavy KK modes induced by the higgs vev has already been taken into account (to all orders in $\ti m_L^2$).   
The diagonalization is achieved  if the profiles solve the bulk equation of motion: 
\beq
\label{e.geom} % gauge equation of motion 
\left (\pa_5^2  + 2 {a' \over a} \pa_5 +  {m_n^2 \over a^2} \right ) f_n(x_5) = 0
\eeq
and satisfy appropriate boundary conditions.  
On the UV brane, in absence of any localized mass  or kinetic terms, these are the Neumann or Dirichlet boundary conditions,  
\beq
\label{e.bcuv}
\pa_5 f_n(0) = 0 \qquad {\rm or} \qquad f_n(0) = 0 
\eeq
On the IR brane we should impose  
\beq
\label{e.bcir}
\pa_5 f_n(L) =   - L a_L^{-2} \ti m_L^2 f_n(L)
\eeq
The Dirichlet boundary conditions can be simulated in the limit $\ti m_L^2 \to \infty$ (we call it the \emph {higgsless limit}). 
The  profiles should  also satisfy the normalization condition 
$\int_0^L |f(y)|^2 = 1$.

The usual procedure is to solve the equations of motion \eref{geom} for some particular background. 
In this paper we show how to obtain results valid for an arbitrary warp factor.   
To proceed, we denote the two independent solutions of \eref{geom} by $C(x_5,m_n)$ and $S(x_5,m_n)$. 
We choose them such that they  satisfy the initial conditions $C(0,m_n) = 1$, $C'(0,m_n) = 0$, $S(0,m_n) = 0$, $S'(0,m_n) = m_n$.  
These functions can be viewed as a warped generalization of the cosines and sines 
(in the flat background $C= \cos(x_5 m_n)$, $S= \sin(x_5 m_n)$). 
Using them, we can write down the profiles in a compact form.  
For example, a profile with the Neumann boundary conditions in the  UV is written as  
$f_n(x_5) = \alpha_n C(x_5,m_n)$,  where $\alpha_n$ is fixed by the normalization condition. 
The spectrum of the KK modes is determined by the IR boundary condition that, in this language, is written as 
$C'(L,m_n) =   - L a_L^{-2} \ti m_L^2 C(L,m_n)$.  
 
Our basic tool will be the expansion of the profiles corresponding to  light fields in powers of $m_n$.
Solving \eref{geom} perturbatively in $m_n$ we can expand the two solutions as 
\bea
\label{e.cssm}
C(x_5,m_n) &=&  1 -  m_n^2 \int_0^{x_5} dy \, y \, a^{-2}(y) + \co(m_n^4)
\nn
S(x_5,m_n) &=&  m_n \int_0^{x_5}dy \, a^{-2}(y) + \co(m_n^3)
\eea 
%\bea
%C(x_5,m_n) &=&  1 -  m_n^2 I_2(x_5) % \int_0^{x_5} (y a^{-2}(y)) 
%+ m_n^4 I_6(x_5) % \int_0^{x_5} a^{-2}(y) \int_0^{y}\int_0^{y'} (y'' a^{-2}(y'')) 
%+ \co(m_n^6)
%\nn
%S(x_5,m_n) &=&  m_n  L^{-1} I_3(x_5) \left \{  1 % L \int_0^{x_5}a^{-2}(y) 
%-  m_n^2 I_5(x_5) % \int_0^{x_5} a^{-2}(y) \int^y\int^{y'} a^{-2}(y'') 
%+ \co(m_n^4) \right \}
%\eea
%The integrals of the warp factor 
%where the integrals $I_k$ of the warp factor are defined in \aref{i}.
We will employ this  expansion for the profiles of the W and Z boson, whose masses are of the order of the electroweak scale.  
This makes sense when the electroweak scale is hierarchically smaller than the KK scale defined\footnote{
This scale is parametrically of the order of the mass of light spin 1 resonances. 
In 5D Minkowski the first KK photon mass is exactly equal to $\mkk$, while in \ads\, it is 
approximately $3/4 \mkk$. }
as 
\beq
\mkk = {\pi \over \int_0^L dy a^{-1}(y)}
\eeq  
In any realistic set-up, a mass gap between the SM gauge fields  and heavy resonances must be large enough to justify this expansion. 
Technically speaking, there are two ways to introduce the mass gap.  
One is to  introduce it by hand %the hierarchy between the electroweak scale and the KK scale, 
by choosing $\ti m_L/\mkk \ll 1$.  
In such a case the electroweak scale is of order $\ti m_L$.   
In our setup the ratio $\ti m_L/\mkk$ can be made arbitrarily small, however as soon as $\mkk \gg 1 \tev$ we face the  hierarchy problem.    
The mass gap may also  exist in the higgsless limit when $\ti m_L \to \infty$.  
In that case the electroweak (lightest resonance) scale and the KK (heavy resonance) scale are related by 
\beq
m_W^2 \sim {1 \over  \int_0^{L} y a^{-2}(y) } \sim {2 \over \pi^2}   {\mkk^2 \over {\cal V } } .
\eeq 
% Changes for the referee 
If the warp factor decreases sharply toward the IR brane the denominator scales linearly with  the size  of the extra dimension $L$.  
To stress this, we  introduced  the {\it volume factor} ${\cal V}$ defined as
\beq
\label{e.vf}
\cv  = L a_L^{-1} \mkk /\pi 
\eeq
To obtain the scale separation between the electroweak and the KK scale we need $\cv$ large enough.     
For  backgrounds that solve the hierarchy problem we indeed expect the volume factor to be large. 
The argument is purely heuristic. 
Typically, solving the hierarchy problem involves generating the huge ratio $\mpl/\tev$ from a moderate number, say, of order $4\pi$.
Moreover, the hierarchy should be generated dynamically by stabilization, thus our moderate number should be somehow correlated with $L$.  
Next, we have  $\mkk \sim a_L k$ where $k \sim a'(L)/a_L$ is the scale that describes how quickly the warp factor changes close to the IR brane.
Then $k L$ is a dimensionless parameter which we may identify with the one that generates hierarchy.     
This is of course not a proof and one can certainly find counterexamples, if the warp factor is complicated enough. 
For simple warp factors, however, the argument works.  
For example, for the \ads\, background the volume factor is $ {\cal V} = \log{a_L^{-1}} \sim \log (\mpl/\mkk) \sim 30$.
% so its inverse provides a perfect expansion parameter. 
Thus, the volume factor is large when the hierarchy is generated by a dynamics that is approximately conformal over a large range of scales.   
On the other hand, in the flat space (which does not address the hierarchy problem) ${\cal V} = 1$.  
%and one could not employ the expansion \erefn{cssm} in the higgsless limit.  
%end of changes for referee   

A similar background independent formalism can be worked out for fermions.
We do not review it here since we limit our study to gauge boson contribution, but see \cite{fa}. 
See also refs. \cite{hisa1,hisa2} for another background independent approach via sum rules. 

%**********************************************************************
\subsection{Effective standard model and dimension-six operators} 
\label{s.d6}
%***********************************************************************

We write the effective low energy theory as 
$\cl_{eff} = \cl_{\mathrm{SM}} +  \cl_{D6}$ 
where $\cl_{\mathrm{SM}}$ is the electroweak part of the SM lagrangian: 
\bea &
\label{e.smlo} % SM lowest order    
\cl_{\mathrm{SM}}  =  
 - {1 \over 2} \tr \{ L_{\mu\nu} L_{\mu\nu} \} -  {1 \over 4 } B_{\mu\nu} B_{\mu\nu} 
+ i \sum_j \ov{\psi_j} \gamma_\mu D_\mu \psi_j
+ |D_\mu H|^2 - V(H)  + {\rm Yukawa} 
\nl
D_\mu \psi_j =
(\pa_\mu  - i  g_L L_\mu^a T^a - i g_Y Y_j B_\mu) \psi_j
\nl
D_\mu H = 
(\pa_\mu - i g_L L_\mu^a T^a  - i g_Y {1 \over 2} B_\mu) H
\eea  
and $\cl_{D6}$ are the dimension-six operators 
\bea 
\label{e.smd6} % standard model dimension six operators
& 
\cl_{D6} = \alpha_T |H^\dagger D_\mu H|^2 + 
\alpha_S (H^\dagger T^a H) W_{\mu\nu}^a B_{\mu\nu}
\nl 
- \left \{  i \beta_j g_L^2 (\ov \psi_j \gamma_\mu t^a \psi_j)  (D_\mu H^\dagger t^a H)
+ i {\gamma_j \over 2} Y_{j} g_Y^2 \ov{\psi_j}\gamma_\mu \psi_j  (D_\mu H^\dagger  H) + \hc \right \} 
\nl + {\rm fermion}^4
% +   (\ov \psi_j \gamma_\mu t^a \psi_k) (H^T t^a D_\mu H) 
\eea
The four-fermion terms will be ignored in the following (in this paper we restrict our discussion to the situations where they do not introduce significant constraints).
There are also other dimension-six operators that can be generated by 5D physics 
(for example,  $(\ov \psi_j \gamma_\mu t^a \psi_k) (H^T t^a D_\mu H)$)
but are ignored here because they do not get large contributions from KK gauge bosons.   
%We assume that the dimensionful coefficients in  $\cl_{D6}$ are much smaller than $1/v^2$.

When the Higgs field acquires the vev we define the photon, W and Z as usual, 
\bea
W_\mu^\pm &=& {1 \over \sqrt 2} (L_\mu^1 \mp i L_\mu^2)
\nn
A_\mu  &=&  {1 \over \sqrt{g_L^2 +  g_Y^2}} \left ( g_Y L_\mu^3 + g_L B_\mu \right ) 
\nn
Z_\mu  &=& {1 \over \sqrt{g_L^2 +  g_Y^2}} \left (g_L L_\mu^3 - g_Y B_\mu \right ) 
\eea 
but their masses and interactions are modified by the dimension six operators. 
The vertex correction $\beta_j$ and $\gamma_j$ modify the interactions of the SM fermions with the W and Z bosons:  
\bea & 
\cl_{eff} \to   {g_L g_Y \over \sqrt{ g_L^2 + g_Y^2}}(t_i^3 + Y_i) \ov \psi_i \gamma_\mu \psi_i A_\mu 
\nl 
+ {g_L \over \sqrt{2}} (1 + m_W^2  \beta_j) \ov{\psi_j} \gamma_\mu t^\pm \psi_j W_\mu^\pm 
\nl
+ {1 \over  \sqrt{g_L^2 + g_Y^2}} 
 (g_L^2(1 + \beta_j  m_Z^2) t_j^3 -  
 g_Y^2 (1 + \gamma_j  m_Z^2) Y_{j} ) \ov{\psi_j} \gamma_\mu \psi_j Z_\mu  
\eea
The Z boson mass is modified by $\alpha_T$ 
\beq
m_W^2 = {g_L^2 v^2 \over 4}  \qquad \qquad 
m_Z^2 =  {(g_L^2 + g_Y^2)  v^2 \over 4}  \left (1 +  {v^2 \over 2}  \alpha_T  \right ) 
\eeq
Finally, $\alpha_S$ mixes the photon and the Z boson, 
\beq
\cl_{D6} \to - {1 \over 4} \alpha_S v^2 L_{\mu\nu}^3 B_{\mu\nu} 
\eeq 

We can adjust the coefficients of the dimension-six operators to match the effective lagrangian obtained by integrating out the KK modes. 
Note that this set of coefficient is redundant: the universal shift of the vertex corrections can be absorbed by redefinitions of the gauge couplings \cite{agdema}. We can  thus shift $\beta_j$ and $\gamma_j$  by $\Delta \beta$ and $\Delta \gamma$ without changing the physical content of the theory, provided  that $\alpha_T$ and $\alpha_S$ are also shifted accordingly: 
\bea 
\label{e.vsts} % vertex S T shift 
\beta_j &\to & \beta_j + \Delta \beta 
\nn
\gamma_j  &\to & \gamma_j + \Delta \gamma   
\nn
\alpha_S &\to & \alpha_S - {g_L g_Y \over 2} \left (  \Delta \beta + \Delta \gamma  \right ) 
\nn 
\alpha_T   &\to & \alpha_T  +  g_Y^2 \Delta  \gamma
\eea
One particular application of this result is when the vertex corrections are universal: 
$\beta_j = \beta$ and $\gamma_j = \gamma$. 
Then, choosing  $\Delta \beta = - \beta$,  $\Delta \gamma = - \gamma$ we can get rid of the vertex corrections, which reemerge as a shift of  $\alpha_T$ and $\alpha_S$. This is the \emph{oblique case}, in which all the corrections from new physics can be parametrized by  $\alpha_T$ and $\alpha_S$.   
Those two are simply related to the  familiar  S  and T  parameters:  
\beq
\label{e.tst} % translation S T 
S  = {8 \pi v^2 \over g_L g_Y}  \alpha_S \qquad  
T =  - {2 \pi v^2 \over e^2} \alpha_T  
\eeq

%%%%%%%%%%%%%%%%%%%%%%%%%%%%%%%%%%%%%%%%%%%%%%%%%
\section{No Custodial}
\label{s.321} 
%%%%%%%%%%%%%%%%%%%%%%%%%%%%%%%%%%%%%%%%%%%%%%%%%

We first consider a 5D model without a custodial symmetry.
The bulk gauge symmetry is  that of the SM, $SU(3)_c \times SU(2)_L \times U(1)_Y$.
The electroweak group is broken to $U(1)_{em}$ by a higgs doublet $H$  localized on the IR brane.
The 5D action for the electroweak gauge bosons and the higgs reads  
\bea & 
\label{e.wsma} % warped SM action   
S= \int d^4 x \int_0^L dx_5 \sqrt{g} \left \{
 -  {1 \over 2} \tr \{ L_{MN} L_{MN} \} -  {1 \over 4} B_{MN} B_{MN} 
\right \} 
\nl
+ \int d^4 x dx_5 \sqrt{g_4} \delta(L) \left \{ |D_\mu H|^2 - V(H)  \right \}  \, , 
\eea
where  $\pa_\mu H = (\pa_\mu - i  \sqrt{L} g_L B_\mu  - i \sqrt{L}  g_Y Y_H B_\mu) H$ and 
$ \sqrt{L}  g_L$, $\sqrt{L}  g_Y$ are the dimensionful gauge couplings of $SU(2)_L \times U(1)_Y$. 
The Higgs field acquires a vev $\la H \ra = (0, \ti v a_L^{-1}/\sqrt 2)^T$. 
This results in  the IR brane mass terms 
\beq
\cl = \delta(L) \left ( {1 \over 2} L {g_L^2 \ti v^2 \over 4 }  |W_\mu^+|^2 
+   {1 \over 2} L {(g_L^2 + g_Y^2) \ti v^2 \over 4}   |Z_\mu|^2 \right ) \, . 
\eeq 
%The boundary masses are related to the Higgs vev by 
%\beq
%\ti m_W^2 = {g_L^2 \ti v^2 \over 4}  
%\qquad 
%\ti m_Z^2 = {(g_L^2 + g_Y^2) \ti v^2 \over 4}  
%\eeq 
On the UV brane we impose the Neumann boundary conditions on the profiles 
\beq
\label{e.wsmbc} % warped SM boundary conditions 
\pa_5 f_n^\gamma(0)= \pa_5 f_n^W(0) = \pa_5 f_n^Z(0) = 0  
\eeq
so that no gauge symmetry gets broken there. 
Using the formalism introduced in \sref{f}, the gauge boson  profiles  satisfying the UV boundary condition can be written as  
\beq
\label{e.321gbc} % 321 gauge boundary conditions 
f_{n}^{W}(x_5) = c_{W,n} C(x_5,m_{W,n})
\quad
f_{n}^{Z}(x_5) = c_{Z,n} C(x_5,m_{Z,n})
\quad
f_{n}^{\gamma}(x_5) = c_{\gamma,n} C(x_5,m_{\gamma,n})
\eeq
On the IR brane, in the presence of the boundary mass terms, the boundary conditions read
\beq
\pa_5 f_n^\gamma(L) = 0
\qquad 
\pa_5 f_n^W(L) = - {L g_L^2 \ti v^2 \over 4 a_L^2}  f_n^W(L)
\qquad 
\pa_5 f_n^Z(L) = - {L (g_L^2 + g_Y^2) \ti v^2 \over 4 a_L^2}  f_n^Z(L)
\eeq 
Those imply the quantization condition for the photon, W and Z mass towers  
\bea  
C'(L,m_{\gamma,n}) & = & 0 
\nn
C'(L,m_{W,n}) + {L g_L^2 \ti v^2 \over 4 a_L^2}  C(L,m_{W,n}) & = & 0   
\nn
C'(L,m_{Z,n}) + {L (g_L^2 + g_Y^2) \ti v^2 \over 4 a_L^2} C(L,m_{Z,n}) & = & 0
\eea

Let us concentrate on the zero modes (we will omit the $n=0$ index).
The photon profile is a constant  
\beq
\label{e.zmpp}%zero mode profile photon
f^\gamma(x_5) = {1 \over \sqrt{L}} \qquad m_\gamma = 0 
\eeq
The W and Z profiles are non-trivial due to the boundary mass terms and their shapes depend on the background geometry. 
However, we dispose of a small parameter -- the ratio of the gauge boson masses to the KK scale -- in which we can expand the deviation of the profile from a constant.  We find 
\bea
\label{e.zmpwz}%zero mode profile WZ
f^W(x_5) &=&  
{1 \over \sqrt{L}} \left (1  +  m_W^2 [I_1(L) - I_2(x_5)] + \co(m_W^4/M_{KK}^4) \right )  
\nn
f^Z(x_5) &=&  
{1 \over \sqrt{L}} \left (1  +  m_Z^2[I_1(L) - I_2(x_5)] + \co(m_Z^4/M_{KK}^4) \right ) 
\eea
where the integrals $I_n$ depend on the warp factor and are defined in \aref{i}.
We can insert this expansion into the boundary conditions \erefn{321gbc}. 
As long as $\ti v$ much smaller then the compactification scale, the gauge boson masses can be perturbatively expanded in $\ti v^2$. 
%\bea
%\label{e.gbm}
%m_W^2 & = &
%\ti m_W^2 \left [  1  + \ti m_W^2 (I_1(L) - I_2(L))  \right ] + \co(\ti m_W^6)
%\nn
%m_Z^2 & = &  
%\ti m_Z^2\left [  1  + 
%\ti m_Z^2 (I_1(L) - I_2(L)) \right ] + \co(\ti m_W^6)  
%\eea
Defining the electroweak scale $v$ by 
\beq
v^2  =  \ti v^2 \left ( 1  + {g_L^2 \ti v^2 \over 4} \left [ I_1(L) - I_2(L) \right ]  \right)
\eeq
we can  write the gauge boson masses as 
\beq
\label{e.wzm}
m_W^2 = {g_L^2  v^2 \over 4} 
\qquad
m_Z^2 = {(g_L^2 + g_Y^2)  v^2 \over 4}  \left (
1 + {g_Y^2 v^2 \over 4} \left [ I_1(L) - I_2(L) \right ]   \right ) \, .
\eeq

We move to discussing the fermionic sector of the model. 
The fermions can be simply realized by assigning a 5D bulk field to each SM fermion.
For example, one quark generation is  contained in the 5D fields
\beq
\ba{rrr} 
q = \bvec u \\ d \evec & \  u^c  \ &  d^c \ 
\\
\bf 2_{1/6} \ &  \bf 1_{2/3} & \bf 1_{-1/3}
\ea 
\eeq  
with the action 
\bea &
S= \int d^4 x \int_0^L dx_5 \sqrt{g} \{ 
 \ov q (i \Gamma_N D_N + M_q) q 
  + \ov{u^c}(i \Gamma_N D_N - M_u) u^c 
  \nl
+ \ov{d^c}(i \Gamma_N D_N - M_d) d^c  \}  
-  \int d^4 x dx_5 \sqrt{g_4} \delta(L) \left ( 
\ti y_u \ov{q_L}  H^\dagger  u_R^c +  \ti y_d  \ov{q_L} H d_R^c + \hc  \right ) 
\eea
For the light generations we  can ignore the mixing between the zero modes and the  heavy KK modes. 
The zero mode profiles are then 
\beq
f_L^q \approx {a^{-2}(x_5) e^{-M_q x_5}
\over \int_0^L a^{-1}(y) e^{-2 M_q y} }
\quad 
f_R^{u^c} \approx {a^{-2}(x_5) e^{-M_u x_5}
\over \int_0^L a^{-1}(y) e^{-2 M_u y} }
\quad
f_R^{d^c} \approx {a^{-2}(x_5) e^{-M_d x_5}
\over \int_0^L a^{-1}(y) e^{-2 M_d y} }
\eeq 
and the quark masses are related to the boundary Yukawa couplings by 
\bea &
m_u^2\approx
{a_L^{-2} e^{-(M_q + M_u)L}   (y_u \ti v)^2 \over 
\int_0^L a^{-1}(y) e^{-2 M_q y} \int_0^L a^{-1}(y) e^{-2 M_u y}
} 
\nl
m_d^2 \approx
{a_L^{-2} e^{-(M_q + M_d)L} (y_d \ti v)^2  \over 
\int_0^L a^{-1}(y) e^{-2 M_q y} \int_0^L a^{-1}(y) e^{-2 M_d y}
} 
\eea

We can now write down the interactions of the light fermions with the light gauge bosons 
\ben
\item Electromagnetic currents: 
\beq
 \cl_{em} = e (t_j^3 + Y_j) \ov \psi_j \gamma_\mu \psi_j A_\mu 
\qquad 
e =  {g_L g_Y \over \sqrt{ g_L^2 + g_Y^2}}
\eeq
\item Charged currents: 
\beq
\label{e.cc}
\cl_{cc} = 
{g_L \over \sqrt{2}} [1 + m_W^2(I_2(L) - J_2(L,M_j)) ]  
\ov \psi_j\gamma_\mu t_j^\pm \psi_j W_\mu^+  + \hc
\eeq
\item Neutral currents: 
\beq
\label{e.nc}
\cl_{ncL}  = 
{g_L^2 t^3_j -  g_Y^2 Y_j  \over \sqrt{g_L^2 + g_Y^2}}
[1 + m_Z^2 (I_2(L) - J_2(L,M_j)) ]
\ov \psi_j \gamma_\mu \psi_j Z_\mu 
\eeq
\een

These gauge interactions and the corrections to the Z boson mass can be reproduced by the SM lagrangian  supplemented by the dimensions six operators defined in \eref{smd6}.
We find the coefficients   
\bea
\label{e.sd6} % standard dimension six
\alpha_T &=&  - {1 \over 2}  g_Y^2 \left ( I_1(L) +  I_2(L) \right )  
\nn
\alpha_S &=&  g_L g_Y I_1(L)
\nn
\beta_j &=&   -  J_2(L,M_j)
\nn
\gamma_j &=&   - J_2(L,M_j)    
\eea

The vertex corrections are in general non-universal, so that the corrections are not oblique. 
However, when the fermions are localized near  the UV brane, the vertex corrections are negligibly small.
This is the typical assumption for the first two generations of the SM fermions.  
In such case the corrections can be treated as oblique and adequately parametrized by the familiar S and T:
\beq
\label{e.sst} % standard S T 
S  = 8 \pi v^2  I_1(L) 
\qquad \qquad   
T = {4 \pi  m_Z^2 \over g_L^{2}} \left ( I_1(L) +  I_2(L) \right )
\eeq
%As discussed at more length in  \aref{i}, 
Because of the volume enhancement, the integral $I_2$ dominates for backgrounds with the large volume factor.
Thus we get an approximate expression for the T parameter:
\beq
\label{e.321t} %  T parameter
T \approx 4 \pi  m_Z^2 g_L^{-2} \int_0^{x_5} y a^{-2}(y) \sim {2 \pi^3 m_Z^2 \over g_L^2 \mkk^2}  {\cal V} 
\eeq  
Here ${\cal V}$ is  the volume factor introduced in  \eref{vf}.    
For  backgrounds that solve the hierarchy problem we expect the volume factor to be large, which would  strongly enhance the contribution to T. 
For example, in the Randall--Sundrum model $ {\cal V}  \sim \log (\mpl/\mkk) \sim 30$ leading to a very strong constraint on the KK scale \cite{cserte}. 
Our results show that the problem persists in any 5D warped model without a custodial symmetry, in which the solution to the hierarchy problem is associated with a moderately large volume factor.

%%%%%%%%%%%%%%%%%%%%%%%%%%%%%%%%%%%%%%%%%%%%%%%%%
\section{Custodial}
\label{s.3221} 
%%%%%%%%%%%%%%%%%%%%%%%%%%%%%%%%%%%%%%%%%%%%%%%%%

The well-known cure for an excessive T parameter is the custodial $SU(2)_R$ symmetry.  
In the context of 5D  theories the custodial symmetry is promoted to a gauge symmetry.  
The hypercharge group is extended to $SU(2)_R \times U(1)_X$ that is broken to $U(1)_Y$ on the UV brane \cite{agdema}.
Thus, the bulk gauge symmetry is   $SU(3)_c \times SU(2)_L \times SU(2)_R \times U(1)_X$. 
%The IR brane should not break the bulk gauge symmetry. 
The IR brane higgs field $\Phi$ is in the  $(2,\ov 2)_0$ representation with respect to $SU(2)_L \times SU(2)_R \times U(1)_X$.
Below we apply our background independent techniques to this class of models. 
 
The 5D action for the (extended) electroweak sector reads 
\bea &
\label{e.wcma} % warped SM action   
S= \int d^4 x \int_0^L dx_5 \sqrt{g} \left \{
-{1 \over 2} \tr \{ L_{MN} L_{MN} \} -{1 \over 2} \tr \{ R_{MN} R_{MN} \} -  {1 \over 4 } X_{MN} X_{MN} 
\right \}
\nl
 \int d^4 x dx_5 \sqrt{g_4} \delta(L) \left ( 
 {1 \over 4} {\rm Tr} |D_\mu \Phi|^2 - V(\Phi)  
\right ) 
\eea 
The higgs field acquires the vev $\la \Phi \ra = {\ti v \over a_L} I_{2 \times 2}$. 
This results in the mass terms 
\beq
\cl_{mass} = {1 \over 8} L \ti v^2 \delta(L) (g_L L_\mu^a - g_R R_\mu^a)^2 
\eeq
that spontaneously break $SU(2)_L \times SU(2)_R$ to $SU(2)_V$ on the IR brane.

The UV  boundary conditions that break $SU(2)_R \times U(1)_X$ down to $U(1)_Y$ impose the following conditions on the KK profiles: 
\bea
\pa_5 f_{L,n}^a(0) &=& 0  \qquad  a = 1,2,3 
\nn
f_{R,n}^i(0) &=& 0 \qquad  i = 1,2 
\nn
s_x \pa_5 f_{R,n}^3(0) + c_x \pa_5 f_{X,n}(0) &=& 0 \qquad  s_x = {g_X \over \sqrt{g_X^2 + g_R^2}}  
\nn
-c_x f_{R,n}^3(0) + s_x f_{X,n}(0) &=& 0 \qquad  c_x = {g_R \over \sqrt{g_X^2 + g_R^2}} 
\eea
The profiles that solve the bulk equations of motion and satisfy the UV boundary conditions in our notation are written as  
\bea
f_{L,n}^a(x_5) &=& \alpha_{L,n}^a C(x_5,m_n)
\nn
f_{R,n}^i(x_5) &=& \alpha_{R,n}^i S(x_5,m_n)
\nn
f_{R,n}^3(x_5) &=& \alpha_{N,n} s_x  C(x_5,m_n) - \alpha_{D,n} \,  c_x S(x_5,m_n)
\nn
f_{X,n}(x_5) &=& \alpha_{N,n} c_x  C(x_5,m_n)+ \alpha_{D,n} \, s_x S(x_5,m_n)
\eea
The linear combination $B_\mu = s_x R_\mu^3 + c_x X_\mu$ survives on the UV brane and its zero mode is identified with the hypercharge gauge boson. 
$B_\mu$ couples to matter with the coupling $g_Y = g_X g_R/\sqrt{g_X^2 + g_R^2}$ and the hypercharge depends on the $SU(2)_R \times U(1)_X$ quantum numbers via $Y = t_R^3 + X$.
    
In the presence of the higgs vev, the IR boundary conditions for the profiles read 
\bea
\label{e.wcmbcir} % warped CM boundary conditions in IR 
\pa_5 f_{X,n}(L) &=& 0 
\nn
 g_R \pa_5 f_{L,n}^a(L) + g_L \pa_5 f_{R,n}^a(L)  & = & 0 
\nn
g_L \pa_5 f_{L,n}^a(L) - g_R \pa_5 f_{R,n}^a(L)  & = & 
-    {1 \over 4} (g_L^2 + g_R^2) a_L^{-2} L \ti v^2 (g_L f_{L,n}^a(L)  - g_R f_{R,n}^a(L)) 
\eea
We can now solve these equations to find the mass eigenstates. 
The spectrum contains a tower of charge $\pm 1$ massive vector bosons whose masses are given by the solutions of the quantization condition:
\bea &
\label{e.cwm} % custodial W mass 
S'(L,m^W_n) C'(L,m^W_n) + \nl
{a_L^{-2}  L \ti v^2 \over 4} \left ( g_L^2 S'(L,m^W_n) C(L,m^W_n) + g_R^2 S(L,m^W_n) C'(L,m^W_n) \right )  = 0
\eea
and a tower of electrically neutral massive vector boson with masses
\bea &
\label{e.czm} % custodial Z mass 
S'(L,m^Z_n) C'(L,m^Z_n) +\nl
   { a_L^{-2} L \ti v^2 \over 4} \left ( g_L^2 S'(L,m^Z_n) C(L,m^Z_n) + g_R^2 S(L,m^Z_n) C'(L,m^Z_n) + m^Z_n a_L^{-2} g_Y^2  \right ) = 0 
\eea
There is also another tower of neutral vector bosons with masses given by 
\beq
C'(L,m_n^\gamma)  = 0 
\eeq  
which always includes the photon solution with $m_\gamma  = 0$.
The lightest solutions of eqs. (\ref{e.cwm}) and (\ref{e.czm}) are identified with the W and Z boson masses. 
  
We turn to  discussing the profiles of the zero mode fields that correspond to the SM gauge bosons. 
In phenomenologically viable  models these profiles can always be expanded in powers of the gauge boson masses over the KK scale.   
To first order\footnote{At the second order there exist $m_W^4/\mkk^4$ terms enhanced by the volume factor $\cv$. Those terms can be neglected if $m_W^2/\mkk^2 < 1/\cv$, which we assume in this section. On the other hand, in the higgsless case $m_W^2 \sim \mkk^2/\cv$ and these terms have to be retained.}
 $m^2/\mkk^2$ we find the SM gauge bosons are embedded into the 5D fields as follows   
\bea
\label{e.ccgp} % custodial charged gauge boson profiles 
L_\mu^i(x,x_5) &\to&  {1 \over \sqrt{L}} \left ( 1 +  m_W^2 [I_1(L) - I_2(x_5)] \right )  W_\mu^i(x)   
\nn
R_\mu^i(x,x_5)  &\to&    {1 \over \sqrt{L}}  m_W^2 {g_R \over g_L}  I_3(x_5) W_\mu^i(x) 
\eea 
\bea
\label{e.cngp} % custodial neutral gauge boson profiles 
L_\mu^3(x,x_5) &\to& {1 \over \sqrt L} ( 
\sin  \theta_W A_\mu(x)  
\nn && +  \cos \theta_W ( 1 + m_Z^2 [I_1(L) - I_2(x_5)]) Z_\mu(x)  ) 
\nn
R_\mu^3(x,x_5) &\to& {s_x \over \sqrt L} ( \cos   \theta_W A_\mu(x) \nn
&& - \sin  \theta_W   
 ( 1 +  m_Z^2 [I_1(L) - I_2(x_5) + I_3(x_5) -{1 \over s_x^2} I_3(x_5) ] ) Z_\mu(x) ) 
\nn
X_\mu(x,x_5) &\to& {c_x \over \sqrt L}  (  \cos \theta_W  A_\mu(x) \nn
&& -  \sin \theta_W   
 ( 1 +  m_Z^2  [I_1(L) - I_2(x_5) +  I_3(x_5) ] ) Z_\mu(x) ) 
\eea
The integrals $I_n$ are defined in \aref{i}.
Inserting this expansion into the 5D action we can  read off the interactions between the SM gauge bosons and fermions.  
In general we should also perform the analogous KK expansion for the fermions.  
For the first two generations it is enough to convolute the gauge profiles with the massless fermionic profile  
\beq
f_j(x_5) = { e^{-M_j x_5} a^{-2}(x_5) \over \left ( \int_0^L e^{-2 M_j y} a^{-1}(y) \right )^{1/2} } 
\eeq
and the corrections are of order $m_j^2/\mkk^2$.
We find
\ben
\item Electromagnetic current:
\beq
 \cl_{em} =  { g_L  g_Y \over \sqrt{ g_L^2 + g_Y^2}} (t_{L,j}^3 + t_{R,j}^3 + X_j) \ov \psi_j \gamma_\mu \psi_j A_\mu 
\eeq 
\item Charged current:
\beq
\label{e.ccc} % custodial charged current 
\cl_{cc} = 
{g_L \over \sqrt{2}} \left (1 + m_W^2 [I_1(L) - J_2(L,M_j)] \right )    
\ov \psi_j \gamma_\mu t_j^{\pm} \psi_j W_\mu^\pm   
\eeq
%Note that the viciousness is the same as that of  $I_n$. 
%
\item Neutral current: 
\bea &
\label{e.cnc} % custodial neutral current 
\cl_{ncL}  = {1  \over \sqrt{ g_L^2 + g_Y^2}}
 \left \{ 
g_L^2 t_{L,j}^3 \left (1 + m_Z^2 [I_1(L) - J_2(L,M_j)] \right )
\bnl
-  g_Y^2 Y_j \left (1 + m_Z^2 [I_1(L) - J_2(L,M_j) + J_3(L,M_j)] \right )    
\bnl 
+  g_R^2 t^3_{R,j}  m_Z^2 J_3(L,M_j) \right \} 
\ov \psi_j \gamma_\mu \psi_j Z_\mu 
\eea
\een
At the zeroth order these are just the SM gauge interactions. 
The corrections of order $m^2/\mkk^2$ depend not only on the bulk mass $M_j$ but also on the  embedding of the SM fermions into $SU(2)_R$ representations.  
There are several choices one can make.  
The simplest is to embed the SM  left doublets into  $SU(2)_R$ singlets and the SM left singlets into $SU(2)_R$ doublets \cite{agdema}. 
For example, the lightest quark generation can be embedded as follows 
\beq
\ba{rrr}
q = \bvec u \\ d \evec 
&
u =  \bvec u^c \\ \ti d^c \evec 
&
d  = \bvec \ti u^c \\  d^c \evec 
\\
\bf (2,1)_{1/6} 
&
\bf (1,2)_{1/6}
& 
\bf (1,2)_{1/6}
\ea
\eeq    
The singlet quarks should be put into  two different bulk multiplet to give mass for both the $u$ and $d$ quarks. 
The tilded fermions can be removed from the low energy spectrum by $SU(2)_R$ breaking boundary conditions on the UV brane. 
Another possibility is to embed the SM left doublets into the bifundamental representation of $SU(2)_L\times SU(2)_R$ \cite{agcoda}. 
Then the SM left singlets should be placed into  $SU(2)_R$ singlets or triplets.
For example \cite{caposa} 
\beq
\label{e.3221fbd} % custodial fermions in bidoublets 
\ba{ccc}
Q = \left [
\ba{cc}
u  & \chi 
\\
d & \ti u 
\ea 
\right ] \ 
& 
u^c 
&
D = \left [
\ba{cc}
{1 \over \sqrt {2}} \ti u^c  & \chi^c 
\\
d^c &  - {1 \over \sqrt {2}} \ti u^c 
\ea 
\right ]  
\\
\ \ \ \bf (2,\ov 2)_{2/3}
& 
\bf (1,1)_{2/3}
&
\ \bf (1,3)_{2/3} 
\ea 
\eeq 
Here $\chi^c$ is an exotic charge $5/3$ quark. Again,  by an appropriate choice of boundary conditions we  can ensure that only the SM quarks show up in the low-energy spectrum.

If the boundary Higgs vev is much smaller than the KK scale we can also solve eqs. (\ref{e.cwm}) and (\ref{e.czm}) perturbatively in  $\ti v^2/ \mkk^2$. 
%To quartic order we find  
%\bea
%\label{e.cwzmv} %custodial W Z masses in powers of v
%m_W^2 &\approx& g_L^2 {\ti v^2 \over 4} \left ( 
%1 +  {\ti v^2 \over 4} \left [ 
%\ti g_L^2 (I_1(L) - I_2(L))   -  g_R^2 I_3(L) \right ]\right)  
%\nn
%m_Z^2 &\approx& ( g_L^2+ g_Y^2) {\ti v^2 \over 4} \left ( 
%1 +  {\ti v^2 \over 4} \left [ 
%g_L^2 (I_1(L) - I_2(L))   -  g_R^2 I_3(L)  
%+ g_Y^2(I_1(L) + I_4(L))  \right ]  \right)  
%\eea
Defining
\beq
\label{e.cfs} % custodial fermi scale 
v^2  = \ti v^2  \left ( 
1 +  {\ti v^2 \over 4} \left [ 
 g_L^2 [I_1(L) - I_2(L)] - g_R^2 I_3(L) \right ]  \right)  + \co(\ti v^6)
\eeq 
the gauge boson masses are given by  
\beq
\label{e.cwzm}
m_W^2 = {g_L^2  v^2 \over 4} 
\qquad
m_Z^2  =  {(g_L^2 + g_Y^2) v^2 \over 4}  \left ( 
1 + {g_Y^2 v^2 \over 4} \left [I_1(L) + I_4(L) \right ] \right ) +  \co(v^6)
\eeq
We have now all the necessary information to read off the coefficient of the dimension six operators defined in \eref{smd6}
\bea
\label{e.cd6} % custodial dimension six
\alpha_T &=&  {1 \over 2}  g_Y^2  \left ( - I_1(L) + I_4(L) \right )
\nn
\alpha_S &=&  g_L  g_Y I_1(L)
\nn
\beta_j &=&  -  J_2(L,M_j)
%  {\int_0^L a^{-1}(y) e^{-2 M_j y} I_2(y) \over \int_0^L a^{-1}(y) e^{- 2 M_j y}}
\nn
\gamma_j &=& -  J_2(L,M_j) +  J_3(L,M_j) - {g_R^2  t_{R,j}^3 \over g_Y^2 Y_j } J_3(L,M_j)
 %{\int_0^L a^{-1}(y) e^{-2 M_j y} I_3(y) \over \int_0^L a^{-1}(y) e^{- 2 M_j y}}
\eea
The vertex corrections are non-universal. They depend on both the bulk fermion masses and the embedding into $SU(2)_L \times SU(2)_R \times U(1)_X$. 
Note that in this case the non-universality persists even when all $M_j$ are equal.
%But, similarly as in the case without the custodial symmetry, 
The vertex corrections can be however safely neglected for fermions localized close to the UV brane.  
In such case the corrections can be adequately parametrized by the  S and T parameters:
\beq
\label{e.cst} % custodial S T 
S  = 8 \pi v^2  I_1(L) 
\qquad \qquad   
T = 4 \pi  m_Z^2 g_L^{-2} \left ( I_1(L) -  I_4(L) \right )
\eeq
The oblique parameters do not depend on the fermionic representation. 
The S parameter is given by exactly the same integral as in the case without the custodial symmetry. 
For the T parameter the custodial symmetry is at work: the combination of the integrals that enters there is suppressed, rather than enhanced, by the volume factor ${\cal V}$. 
Thus the tree-level T parameter ends up being tiny for the backgrounds with a large volume factor.

For the third generation fermions we cannot assume that their profiles are localized in the UV since some relevant coupling to the IR brane is needed to give mass to the top quark. 
Thus their interaction vertices with the SM gauge bosons can receive sizable corrections. 
Parameterizing the Z-boson vertex as 
\beq
\cl_{Zjj} = 
{g_L^2 t_{L,j}^3  -  g_Y^2 Y_j   \over \sqrt{g_L^2 + g_Y^2}}  \left ( 
 1  + \delta g_{Zjj} \right )  \ov{\psi_j} \gamma_\mu \psi_j Z_\mu
\eeq
we find the corrections from the gauge bosons exchange\footnote{%
For the third generation there can also be corrections of order $m_T^2/\mkk^2$ from a mixing with the fermionic KK modes. 
Those depend on a precise realization of the fermionic sector and are not discussed here.   
}
\beq
\label{e.cvc} % custodial vertex correction
\delta g_{Zjj}  = 
m_Z^2 (J_3(L,M_j)  - J_2(L,M_j))  
-  m_Z^2 J_3(L,M_j)  { g_L^2 t_{L,j}^3  -   g_R^2  t_{R,j}^3 
 \over   g_L^2 t_{L,j}^3  -  g_Y^2 Y_j }
\eeq
The second term, if not suppressed by the bulk mass exponential, is  volume enhanced and typically dominates the vertex correction:
\beq
\label{e.cvc2} % custodial vertex correction
\delta g_{Zjj}  \approx 
-  { g_L^2 t_{L,j}^3  -   g_R^2  t_{R,j}^3 
 \over   g_L^2 t_{L,j}^3  -  g_Y^2 Y_j }
 m_Z^2 L     {\int_0^L a^{-1}(y) e^{-2 M_j y}  \int_0^{y}  a^{-2}(y') \over
  \int_0^L a^{-1}(y) e^{- 2 M_j y}} 
\eeq 
If $e^{-2 M_j y} < a^{-1}(y)$ close to the IR brane this expression scales linearly with $L$, so that 
$\delta g_{Zjj} \sim {\cal V} m_Z^2/\mkk^2$, similarly as T in the absence of the custodial symmetry.
This can  be dangerous for the  $Z b_L \bar{b}_L$  vertex that is well constrained by experiment,
$\delta_{Z b_L \bar{b}_L} < 0.0025$.
Indeed, in some cases, e.g. for the pseudo-goldstone higgs, this provides the tightest constraint on the parameter space \cite{agco}.   
One can however introduce a symmetry that allows to keep this vertex under control \cite{agcoda} (for another approach, see \cite{djmori}).
As can be seen  from \eref{cvc}, we need a $LR$ parity symmetry that sets $g_L = g_R$. 
Then the second term vanishes  if $b_L$ has $t_L^3 = t_R^3$. 
This is possible if the third generation left doublet originates from  the $(2,\ov 2)$ representation of  $SU(2)_L \times SU(2)_R$.
For example, the embedding in \eref{3221fbd} satisfies this requirement, although there we need additional multiplets in the $(3,1)_{2/3}$ representation to keep the LR parity stable.  
The symmetry advocated in \cite{agcoda} also protects the vertex against order $m_T^2/\mkk^2$ corrections from a mixing of the bottom quark with the fermionic KK modes.   
When the volume enhanced contribution is canceled there still remains an order $m_Z^2/\mkk^2$ contribution from the first term in \eref{cvc}. 
However, it always represents a weaker constraint on the KK scale than that from the S parameter.

%%%%%%%%%%%%%%%%%%%%%%%%%%%%%%%%%%%%%%%%%%%%%%%%%
\section{Higgsless}
\label{s.hl} 
%%%%%%%%%%%%%%%%%%%%%%%%%%%%%%%%%%%%%%%%%%%%%%%%%

The study we performed is slightly modified in the higgsless limit $\ti v \to \infty$. 
Below we consider the custodially symmetric model of \sref{3221} (that of \sref{321} is totally unrealistic in this limit, as it predicts $m_W = m_Z$).  
We cannot, of course, expand in powers of $\ti v/\mkk$ anymore.
Thus, \eref{cfs}  and the following expression for the gauge boson masses are no longer valid. 
On the other hand, the ratio of the gauge boson masses to the KK scale remains a perfect expansion parameter in any realistic setup (including the \ads\, background).
Nevertheless, there is one qualitative change. 
Before, $m_W/\mkk$ was a tunable parameter controlled by $\ti v$.
In the higgsless limit $m_W$ is intimately tied to $\mkk$. 
The precise relation depends on the background geometry and it turns out to be of the form  
$m_W^2 \sim \mkk^2/\cv$.
%This has one practical consequence for expanding the gauge boson profiles. 

Let us first discuss the SM gauge boson masses in a quantitative way. 
Taking the  $\ti v \to \infty$ limit of \eref{cwm} and \eref{czm} we get the quantization conditions   
\bea 
\label{e.hwzm} % higgless W Z  mass  
(g_L^2 + g_R^2) S(L,m_{W}) C'(L,m_{W}) + m_{W} a_L^{-2} g_L^2  &=& 0 
\nn
(g_L^2 + g_R^2)  S(L,m_{Z}) C'(L,m_{Z}) + m_{Z} a_L^{-2} (g_L^2 + g_Y^2)  &=& 0 
\eea
The lightest solution is identified with the SM gauge bosons.
% and  we assume $m_W,m_Z \ll \mkk$ in the following. 
Expanding in $m_{W,Z}/\mkk$ we find  
\bea
m_W^2 & = & {g_L^2 v^2 \over 4}
\nn
m_Z^2 &=&
{(g_L^2 + g_Y^2 ) v^2 \over 4}
\left [
1 +  g_Y^2 { v^2 \over 4} 
\left ( I_1(L) + I_5(L) \right )
\right ] + \co(v^6)
\eea
where we defined 
\beq
\label{e.hfs} % higgsless fermi scale
v^2 =    {4 \over (g_L^2 + g_R^2) L \int_0^L a^{-2}(y)}  \left [
1 +  g_L^2 { v^2 \over 4}  \left ( I_1(L) + I_5(L) \right ) \right ]
\eeq
As advocated, the gap between the electroweak and the resonance scales is controlled by the volume factor,   $m_W^2 \approx \mkk^2/(\pi^2 \cv)$. 
In the following we assume that the 5D background is such that the gap is large enough to allow for expansion in $m_W^2/ \mkk^2$.

The SM gauge boson profiles are those of \eref{ccgp} and \eref{cngp} with 
$I_1(L) \to (I_4(L)+I_5(L))/2$. 
\footnote{The normalization factor changes because there exist 
$\cv m_W^4/\mkk^4$ terms which are of the same order as $m_W^2/\mkk^2$ terms in the higgsless case. For this reason the expression for S is different than in the case with a higgs, where these higher-order terms were neglected.}
From this, we find the coefficients of the dimension-six operators 
\bea
\label{e.hd6} % higssless dimension six
\alpha_T &=&   0 % + \co \left ( {m_W^2 \over \cv \mkk^2} \right )
\nn
\alpha_S &=&  {g_L  g_Y \over 2} \left ( I_4(L) + I_5(L)  \right )
\nn
\beta_j &=&  -  J_2(L,M_j)
\nn
\gamma_j &=& -  J_2(L,M_j) +  J_3(L,M_j) - {g_R^2  t_{R,j}^3 \over g_Y^2 Y_j } J_3(L,M_j)
\eea
where in the oblique parameters we dropped all terms suppressed by the volume factor.  
The T parameter vanishes at this order thanks to the custodial symmetry.
We can write the S parameter as\footnote{%
This expression is consistent with the one derived for general metrics in \cite{hisa1}.}  
\beq
\label{e.hs}
S = 4 \pi v^2 \left ( I_4(L) + I_5(L) \right ) \sim {12 \pi \over \cv (g_L^2 + g_R^2)} 
\eeq 
We would need a large volume factor to suppress the gauge contribution to S. 
Recall that in \ads\, $\cv \sim 30$, which is not enough.
Note that a volume factor large enough to make $S < 0.3$ would also make the KK scale heavier than 1 TeV.
It seems unlikely that the longitudinal WW scattering amplitude could be unitarized by such heavy resonances.
% and so the KK picture could not be valid.  
Therefore the only way to bring $S$ down to acceptable level is by a suitable choice of fermionic profiles. 
More precisely, one assumes that the fermion bulk masses corresponding to the light SM generations are such that the integrals $J_2$ and $J_3$ are not suppressed. 
Moreover, the bulk masses for the different light SM fields should be almost equal, 
$M_j \approx M_{\rm ref}$. 
By \eref{vsts}, choosing $\Delta \beta = J_2$, $\Delta \gamma = J_2 -  J_3$, 
the universal part of the vertex corrections can be traded for a shift in S:
\beq
\Delta S = - 8 \pi v^2 \left (J_2(L,M_{\rm ref}) - {1 \over 2 } J_3(L,M_{\rm ref}) \right ) 
\eeq   
Since the shift is negative, we can cancel the  positive gauge contribution in \eref{hs}  
(in \ads, the cancellation occurs for $M \sim 0.47 k$). 
Note however that there will still remain sizable non-universal vertex corrections, 
$\gamma_j =  - (g_R^2  t_{R,j}^3/g_Y^2 Y_j) J_3(L,M_j)$, which depend on the embedding of the SM fields into $SU(2)_R$.
Thus constraints based just on the oblique parameters can be misleading and one would need a more refined fit as in \cite{caposa2}.   

%%%%%%%%%%%%%%%%%%%%%%%%%%%%%%%%%%%%%%%%%%%
\section{Conclusions}
%%%%%%%%%%%%%%%%%%%%%%%%%%%%%%%%%%%%%%%%%%%%

In this paper we have studied  different models for electroweak breaking based on an extra dimension with a completely general warp factor. Under very broad conditions one can obtain useful expressions for the spectrum of the model and one can then match to a 4D theory with only light particles plus higher dimensional operators. This operators give contributions to the electroweak observables $S,T$, three main conclusions can be drawn:

\begin{itemize} 
\item
If the model is only based on the SM gauge groups there are very large contributions to the $T$ parameter unless the KK scale is very large.
\item
If the model has $SU(2)_R$ as a symmetry in the bulk and then $T$ is under control. As long as there is a higgs in the theory then the model has no problems with EW observables for KK masses of a few TeV.
\item
If we go to the higgsless limit then the $S$ parameter in general grows and some careful choice of fermionic parameters are needed to ensure that the model passes the EW observables test. It is important to note that in this kind of models there is no free parameter to tune in order to reduce $S$ since the KK masses are closely related to the EW scale 
\end{itemize}

All of the above conclusions are \emph{independent} of the geometry of the extra dimension so our conclusions are general. 

Once this models are in agreement with present day bounds one could study the different experimental signatures and resonances that can be produced at LHC. In general the KK modes for gauge bosons tend to be too heavy $\sim 3$ TeV.  More promising signal comes from some light fermionic resonance that can appear when the extra symmetry to cancel $Zb\bar{b}$ couplings is implemented\cite{caposa}.
Another possibility is to study the signal of KK gluons that can be detected even when they are as heavy as 4-5 TeV \cite{kkgluons}. 
 We postpone a detailed study of these signatures until future publications.
%%%%%%%%%%%%%%%%%%%%%%%%%%%%%%%%%%%%%%%%%    

%%%%%%%%%%%%%%%%%%%%%%%%%%%%%%%%%%%%%%%%%
\section*{Acknowledgements}
%%%%%%%%%%%%%%%%%%%%%%%%%%%%%%%%%%%%%%%%%%

AF is partially supported by the European Community Contract MRTN-CT-2004-503369 for the years 2004--2008 
and by  the MEiN grant 1 P03B 099 29 for the years 2005--2007. The authors want to express gratitude to Manuel P\'erez-Victoria for some useful discussions. 

\renewcommand{\thesection}{Appendix \Alph{section}} 
\renewcommand{\theequation}{\Alph{section}.\arabic{equation}} 
\setcounter{section}{0} 
\setcounter{equation}{0} 

%%%%%%%%%%%%%%%%%%%%%%%%%%%%%%%%%%%%%%%%%%%%%%%%%%%%%%%%%%%%%%%%%%%%%%%%%%%%% 
\section{Integrals} 
\label{a.i} 
%%%%%%%%%%%%%%%%%%%%%%%%%%%%%%%%%%%%%%%%%%%%%%%%%%%%%%%%%%%%%%%%%%%%%%%%%%%%%% 

The profiles and the masses of the SM gauge bosons depend on certain integrals of the warp factor. 
Here is the complete list relevant to us:  
\bea
I_1(x_5) &=& L^{-1} \int_0^{x_5} \int_0^y y' a^{-2}(y') 
\nn
I_2(x_5) & = &   \int_0^{x_5} y a^{-2}(y)
\nn
I_3(x_5) &=&  L \int_0^{x_5}  a^{-2}(y)
\nn
I_4(x_5) &=&  \int_0^{x_5} \int_0^y a^{-2}(y')
\nn
I_5(x_5) &=& { \int_0^{x_5}  a^{-2}(y) \int_0^{y} \int_0^{y'} a^{-2}(y'') \over 
\int_0^{L}  a^{-2}(y) } 
\eea
Notice that $I_4(L) = I_3(L) - I_2(L)$. 
All these integrals have dimension $[{\rm mass}]^{-2}$,  therefore $I_n(L)$ is expected to be of order $1/\mkk^2$.
%Changes for the referee
However $I_2(L)$ and $I_3(L)$ can be parametrically enhanced when the volume factor defined in \eref{vf} is large. 
The argument goes as follows. 
If the warp factor decreases sharply toward the IR brane, we get $\int_0^{L}  a^{-2} \sim a_L^{-2}/k$ and also $\mkk \sim a_L k$.
Here, $k \sim a'(L)/a_L$ is the scale that describes how quickly the warp factor changes close to the IR brane.  
Thus $I_3(L) \sim k L /\mkk^2 = \cv/\mkk^2$. Similarly $I_2(L) \sim \cv/\mkk^2$ as the integral is dominated by $y \sim L$. 
Therefore, those electroweak parameters that depend on $I_2(L)$ and $I_3(L)$ end up being larger than the naive estimate $v^2/\mkk^2$, and the constraints on the resonance scale become more stringent. 
Using analogous estimates, the remaining integrals: $I_1(L)$, $I_4(L)$ and $I_5(L)$ are $\sim 1/\mkk^2$, with no volume enhancement. 
Note also that $I_4(L) - I_1(L) =  \int \int (1 - y/L)a^{-2}$ is suppressed, because the integrand vanishes at $y = L$. 
%End of changes for referee 
% these two integrals scale linearly with $L$: they become \emph{volume enhanced}.  

As an example consider the familiar \ads background corresponding to $a(x_5) = \exp(- k x_5 )$. 
For $a_L \ll 1 $ the resonance scale is, $\mkk \approx \pi k a_L$ and the volume factor 
$\cv = k L$ 
We find the integrals 
\beq
I_1(L) \approx {\pi^2 \over 4 \mkk^2} 
\left [1 - {1 \over k L }  \right ]  
\qquad
I_4(L) \approx {\pi^2 \over 4 \mkk^2} 
\qquad
I_5(L) \approx {\pi^2 \over 8 \mkk^2} 
\eeq 
\beq 
I_2(L) \approx {\pi^2 \over 4 \mkk^2} 
\left [2 k L  - 1  \right ]  
\qquad 
I_3(L) \approx {\pi^2 \over 4 \mkk^2}  2 k L   
\eeq 
We can see that, indeed, the integrals in the second line are enhanced by the volume factor, which is of order 
$\log(\mpl/\tev) \sim 30$ in the Randall-Sundrum setup.    

Another, less known example is when the scale factor in 5D varies according to a power law:  
$a(x_5) = \left (1 - {k x_5\over \gamma-1} \right)^\gamma$. 
The parameter $1/\gamma$ allows to describe a departure from conformal symmetry; in the conformal limit 
$\gamma \to \infty$ we are back to   \ads.  
For $a_L \ll 1$ and $\gamma \gg 1$ the resonance scale is  $\mkk \approx \pi k a_L^{1 - 1/\gamma}$ and the volume factor 
$k L a_L^{- 1/\gamma}$.
We obtain:
\bea 
I_1(L) &\approx& {\pi^2 \over 2 \mkk^2}
{\gamma - 1 \over 2 \gamma - 3} 
\left [ 1 - {2 (\gamma-1) \over k L (2 \gamma - 1) }  \right ]  
\nn
I_4(L) &\approx& {\pi^2 \over 2\mkk^2} {\gamma - 1 \over 2 \gamma-1}  
\nn
I_5(L) &\approx& {\pi^2 \over 2 \mkk^2} {\gamma - 1 \over 4 \gamma - 3}  
\eea 
\bea 
I_2(L) &\approx& {\pi^2 \over 2 \mkk^2}  k L a_L^{-1/\gamma}
\left [ 1 - {\gamma-1\over k L (2 \gamma-1)}  \right ]  
\nn 
I_3(L) &\approx& {\pi^2 \over 2 \mkk^2}  k L  a_L^{-1/\gamma}    {2 (\gamma - 1) \over 2 \gamma-1}  
\eea 
We again observe the volume factor popping out in $I_2(L)$ and $I_3(L)$. 
As we move $\gamma$ away from the conformal limit the effect of volume enhancement becomes more dramatic: $I_2(L)$ and $I_3(L)$ grow as a power of the large number $1/a_L$ (rather than a logarithm as in  \ads).  
The remaining integrals, those that are not volume enhanced, are of order $1/\mkk^2$ and weakly depend on the shape  of the warp factor.
In consequence, we cannot significantly reduce the $S$ parameter by varying $\gamma$ 
 ($\gamma$ too close to 1 is not attractive as we need to fine-tune $L$ to generate large hierarchy).     

In the flat space, where there is no hierarchy of scales, all the integrals are of the same order, 
\beq
6 I_1(L) = 2 I_2(L) = I_3(L) = 2 I_4(L) = 6 I_5(L) = {\pi^2 \over \mkk^2}
\eeq  

In order to describe the SM gauge interactions with fermion we define another class of integrals:
\beq
J_n(L, M) =  {\int_0^L a^{-1}(y) e^{-2 M y} I_n(y) \over \int_0^L a^{-1}(y) e^{- 2 M y}} 
\eeq
that depends on the bulk fermion masses.

If the bulk mass is large, such that  
$ a^{-1}(y) e^{- 2 M y} \ll 1$ close to the IR brane, 
then those integrals are suppressed wrt to $1/\mkk^2$. 
For a small bulk mass, however, $J_n$ is of the same order as $I_n$.  

Consider \ads\, once again and parametrize $M = c k$. 
For $c \gg 3/2$ we find $J_n \sim \mkk^{-2} a_L^2$, a Planck scale suppressed result.   
For $1/2 \ll c \ll 3/2$ $J_n \sim \mkk^{-2} a_L^{2c -1}$, still suppressed by an intermediate scale. 
But for $c \ll 1/2$  the suppression is gone and $J_n \sim \mkk^{-2}$. 
In particular 
\beq
J_2(L,c k)  \approx {\pi^2 \over 4 \mkk^2} 
{ (1-2 c) ( 2 k L (3-2 c) - 2 c + 5) 
\over 
(3-2 c)^2 
}
\qquad 
J_3(L,c k)  \approx {\pi^2 \over 4 \mkk^2}  2 k L  
{ 1-2 c
\over 
3-2 c 
}
\eeq 
are volume enhanced, just like $I_2(L)$ and $I_3(L)$. 
But note that $J_3 - J_2$  is safe. 
For the crossover value of the bulk  mass, $c = 1/2$, we obtain
\beq
J_2(L,k/2)  \approx {\pi^2 \over 4 \mkk^2} \left ( 1 - {1 \over k L} \right ) 
\qquad 
J_3(L,k/2)  \approx {\pi^2 \over 4 \mkk^2} 
\eeq 
so that there is no volume enhancement yet.

%*************************************************************************

\end{document}